\documentclass[12pt]{book}

\usepackage[dvips]{graphicx,color}
\usepackage{makeidx,tsukuba}

\makeauthorindex
\makeindex

\begin{document}

\BookTitle{\itshape The 28th International Cosmic Ray Conference}
\CopyRight{\copyright 2003 by Universal Academy Press, Inc.}
\pagenumbering{arabic}

\chapter{
Expected Sensitivity of ARGO-YBJ to Detect Point Gamma-Ray Sources}

\author{%
Silvia Vernetto,$^1$ Carla Bleve,$^2$ Tristano Di Girolamo,$^3$ Giuseppe Di Sciascio,
$^3$ and Daniele Martello$^2$ for the ARGO-YBJ collaboration[6]\\
{\it (1) IFSI-CNR and INFN, Torino, Italy\\
(2) Dipartimento di Fisica Universita' di Lecce and INFN, Lecce, Italy\\
(3) Dipartimento di Fisica Universita' di Napoli and INFN, Napoli, Italy}\\
}

\section*{Abstract}
ARGO-YBJ is a ``full coverage'' air shower detector 
currently under construction
at the Yangbajing Laboratory (4300 m a.s.l., Tibet, China).
First data obtained with a subset of the apparatus will be available
in summer 2003 while the full detector operation is expected in 2005.
One of the main aims of ARGO-YBJ is the observation of gamma-ray sources, at
an energy threshold of a few hundreds GeV.
In this paper we present the expected sensitivity to detect point gamma ray
sources, with particular attention to the Crab Nebula. According to our 
simulations a Crab-like signal could be detected in one year of operation
with a statistical significance of $\sim$10 
standard deviations, without any gamma/hadron discrimination.

\section{The ARGO-YBJ experiment}

The ARGO-YBJ experiment is a ``full coverage'' air shower detector,
optimized to work at a primary energy threshold of a few hundreds GeV.

It consists of a 74 $\times$ 78 m$^2$ ``carpet''
realized with a single layer of Resistive Plate Counters (RPCs),
surrounded by a partially instrumented `` guard ring'', 
for a total active area of 6400 m$^2$. 
The detector is divided into 18480 basic elements, the ``pads'',
of dimensions 56 $\times$ 62 cm$^2$, providing
the space-time pattern of the shower front.
The detector is covered by a 0.5 cm thick layer of lead,
in order to convert a fraction of the secondary gamma rays 
in charged particles, 
and to reduce the time spread of the shower particles. A detailed
description of the experiment is given in [6].
 
ARGO-YBJ is presently under construction 
at the  Yangbajing High Altitude Cosmic Ray Laboratory, in Tibet (China)
at 4300 m above the sea level. 
The full apparatus will be ready in 2005. Presently (spring 2003) a subset
of the detector (about 1650 
m$^2$ of RPCs, corresponding to 28$\%$ of the central carpet) 
has been already installed and it is ready to take the first data. 

One of the main goals of the experiment is the detection of gamma rays from
galactic and extragalactic sources.
The extreme altitude of the detector and the use of
a large full coverage layer of counters, allow the operation at energies
that are typical of the Cerenkov technique.
However, unlike Cerenkov telescopes, 
ARGO-YBJ has a large field of view ($\sim$ 2 sr) and a duty 
cycle $\sim$100$\%$, allowing the simultaneous observation 
of a large fraction of the sky
and making easier the detection of previously unknown gamma ray emitters.

In this paper we evaluate the sensitivity of ARGO-YBJ to detect point 
gamma ray sources, in particular the Crab Nebula, considered the
``standard candle'' for gamma ray astronomy.

\section{Crab Nebula and background event rates}

The daily rate of events from the source has been evaluated by simulating
a gamma ray flux on the top of the atmosphere 
according to the Crab Nebula spectrum
dN/dE = 3.2 10$^{-7}$ E$^{-2.49}$ $\gamma$ m$^{-2}$ s$^{-1}$ TeV$^{-1}$,
measured by the Whipple collaboration [4]. 
The gamma rays have been simulated at
different zenith angles, following the daily path of the source in the sky.
At the Yangbajing site (latitude = 30$^{\circ}$ N) the Crab Nebula culminates
at zenith angle $\theta$ = 8.1$^{\circ}$. We ``followed'' the
source when it was at $\theta\le30^{\circ}$, 
for a total observation time of 4.3 hours per day.

To evaluate the background rate due to cosmic rays we have taken into
account the proton and Helium fluxes on the top of the atmosphere,
according to the spectra dN/dE = 8.98 10$^{-2}$ E$^{-2.74}$ 
protons m$^{-2}$ s$^{-1}$ sr$^{-1}$ TeV$^{-1}$ 
and dN/dE = 7.01 10$^{-2}$ E$^{-2.64}$ Helium nuclei 
m$^{-2}$ s$^{-1}$ sr$^{-1}$ TeV$^{-1}$ [2].
In a first approximation the contribution of heavier nuclei can be neglected.

The showers development in the atmosphere has been
simulated by means of the Corsika code [3]. 
The response of the detector has been studied
by using a GEANT3-based code, that gives 
position and time of all the fired pads
for every shower hitting the detector.

If a shower gives a number of fired pads 
$N_{pad} \ge$50, the position of the core is reconstructed 
and the arrival direction is evaluated by a space-time conical fit of 
the shower front[1,5]. 
Only the events whose reconstructed core falls inside a ``fiducial area''
$A_{fid}$ = 80 $\times$ 80 m$^2$ (centered on the detector) are
considered in the analysis, since the determination of the 
arrival direction is less accurate increasing the distance of the shower core
from the center of the apparatus.

For the events with a number of fired pads $30 \le N_{pad}<$ 50 
the core reconstruction is less reliable, hence a simple planar fit of the
shower front is performed and no selection with the fiducial area is done.

Concerning the determination of the arrival direction of gamma ray showers, 
we found that the angular resolution of the detector 
strongly depends on the number
of fired pads $N_{pad}$, while it is almost independent of the energy and
zenith angle of the primary gamma ray, 
at least in the range of energies and angles considered here.
For this reason the events have been divided into ``classes'' 
defined by ranges of multiplicity $N_{pad}$,
and for each class a different angular resolution has been adopted.
In any class with $N_{pad}\ge$150 the distribution of the angle
$\alpha$ between the true direction (given by the simulation program) 
and the reconstructed one, is well described by a gaussian
distribution dN/d$\alpha$ $\propto$ e$^{-0.5 \alpha^2/\sigma^2}sin\alpha$,
where the parameter $\sigma$ is defined as the angular resolution.
In this case for our ``observations'' we used 
the classical circular window of semi-aperture 
$\psi$ = 1.58 $\sigma$, that contains 71.5$\%$ of the source events and
maximizes the signal to noise ratio.
For $N_{pad}<150$, the angle $\alpha$ distribution is 
not perfectly gaussian and the opening angle that maximizes 
the signal to noise ratio has been separately evaluated for each class.
Fig.1 shows the values obtained for the opening angle $\psi$, 
in different $N_{pad}$ ranges. 

\begin{figure}[t]
\vfill \begin{minipage}[t]{.47\linewidth}
  \begin{center}
    \includegraphics[height=15.3pc]{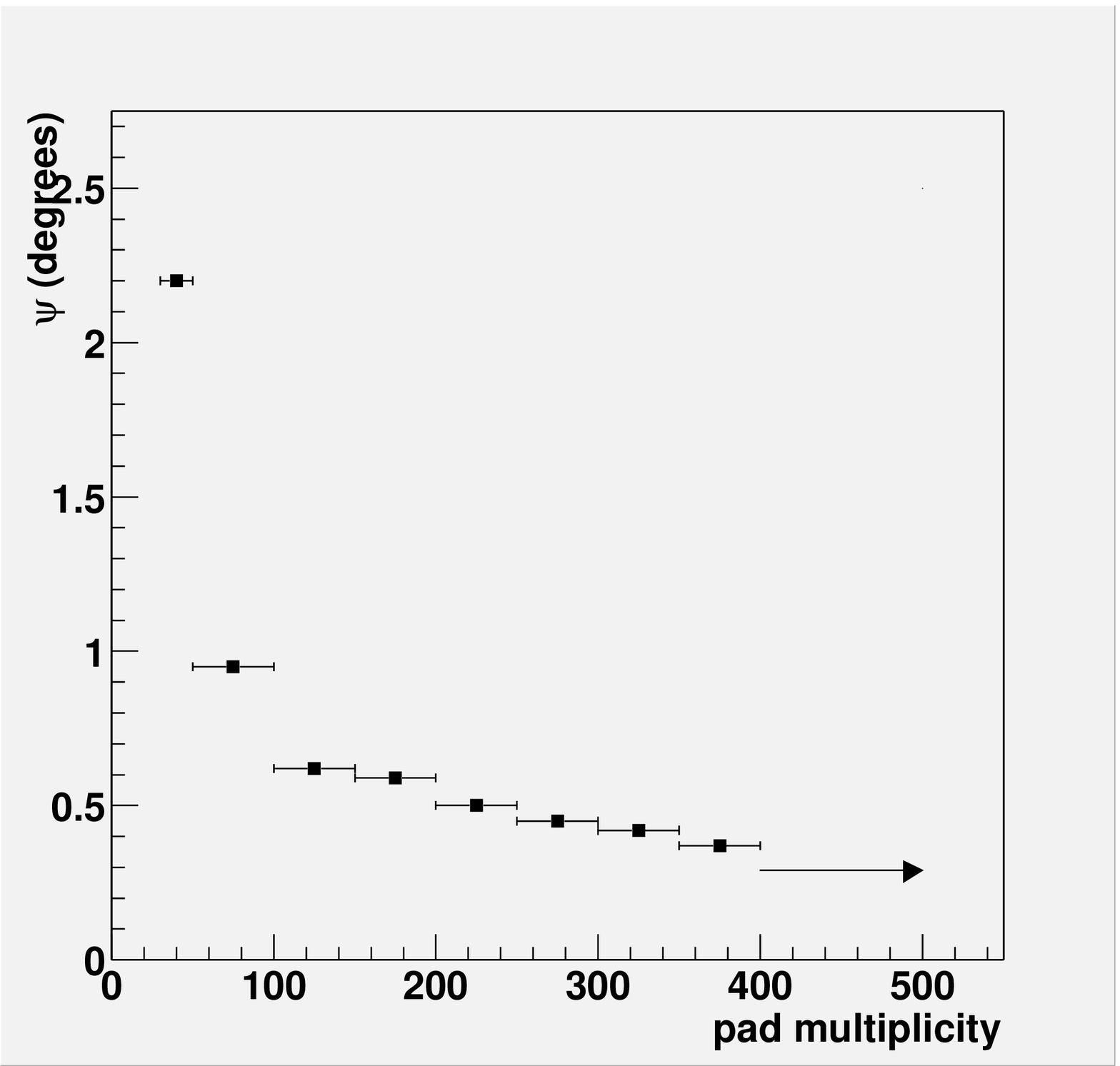}
  \end{center}
  \vspace{-0.5pc}
    \caption{Opening angle of the circular window around the Crab Nebula
position in the sky.}
   \label{llf1_llf2}
\end{minipage}\hfill
\hspace{-0.5cm}
\begin{minipage}[t]{.47\linewidth}
  \begin{center}
    \includegraphics[height=15.3pc]{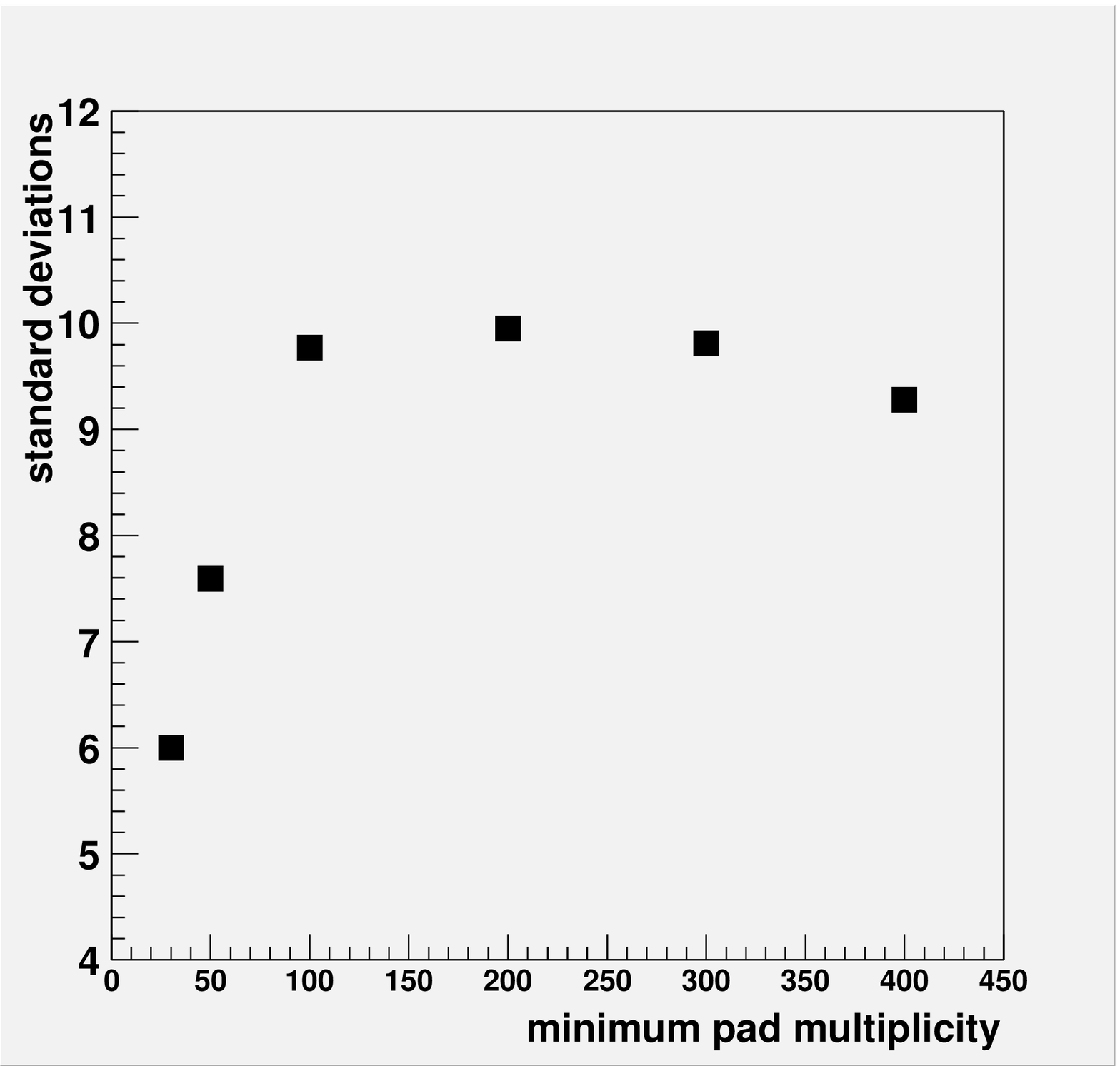}
  \end{center}
  \vspace{-0.5pc}
    \caption{Significance of the Crab Nebula signal after one year of
observation, in terms of standard deviations.}
   \label{finout}
\end{minipage}\hfill
\end{figure}

\section {Results and discussion}

\begin{table}[t]
 \caption{Expected event rates from the Crab Nebula and
from the background in different $N_{pad}$
intervals. The last column reports the median energy of gamma rays.}
\begin{center}
\begin{tabular}{l|cccc}
\hline
$N_{pad}$ & $\gamma$ rays (ev/day) & Prot (ev/day) & He (ev/day) 
& $E_{med}$ (TeV)\\
\hline
30-50       & 58.4 & 5.72 $10^4$ & 1.54 $10^4$ & 0.55 \\
50-100      & 16.5 & 3.73 $10^3$ & 1.12 $10^3$ & 0.60 \\
100-150     & 4.60 & 286         & 87.0        & 0.95 \\
150-200     & 2.42 & 102         & 33.1        & 1.30 \\
200-300     & 2.28 & 56.9        & 21.0        & 1.75 \\
300-400     & 1.09 & 16.6        & 5.8         & 2.32 \\
$>$ 400     & 2.50 & 19.1        & 7.4         & 5.10 \\
\hline
\end{tabular}
\end{center}
\end{table}

Table 1 reports the number of events expected in one 
observation day respectively from the Crab Nebula and from the 
cosmic rays background (protons and Helium nuclei)
for different $N_{pad}$ intervals, using the corresponding observational
windows discussed in the previous section. 
The median energy $E_{med}$ of gamma rays in each multiplicity range 
is also reported.

Adding the contributions of different intervals one 
obtains the daily rate of events with $N_{pad}$ larger than a given
threshold $N_{min}$.
Fig. 2 gives the number of standard deviations (s.d.) of the Crab signal
expected after one year of observation, as a function of $N_{min}$.
According to our calculations,
the signal significance is equal to 6 s.d. for  $N_{min}\ge 30$ 
(corresponding to a primary energy about E $>$ 500 GeV)
then it rapidly increases with $N_{min}$, reaching 
$\sim$10 s.d. for $N_{min}\ge 100$ (about E $>$ 1 TeV) 
and finally slowly decreases for $N_{min}\ge 300$. 

In other words, ARGO-YBJ can observe in one year a source (with a Crab-like
energy spectrum) of intensity equal to 0.7 (0.4) Crab units, 
at energies about $E>$ 0.5 (1.0) TeV,
with a significance of 4 standard deviations.

We expect a certain improvement 
in the sensitivity, particularly in the low energy region, 
by adopting special reconstruction techniques and selection criteria 
for low multiplicity events, now under investigation.

It should be noted
that these results have been obtained without any gamma/hadron
discrimination.
A further increase of the sensitivity could occur performing a
rejection of a large fraction of background events, 
thanks to the different pattern of the shower front particles   
in hadronic and gamma ray induced showers.
Preliminary results on the possibility to separate
background from gamma ray showers  
are very promising and detailed studies are in progress.

\section{References}

\re
1.\ Di Sciascio G.\ et al.\ 2003, Proc XXVIII ICRC
\re
2.\ Gaisser T.K.\ et al.\ 2001, Proc XXVII ICRC, 1643
\re
3.\ Heck D.\ et al.\ 1998, Report FZKA 6019 Forschungszentrum Karlsruhe
\re
4.\ Hillas A.M.\ et al.\ 1998, ApJ 503, 744
\re
5.\ Martello D.\ et al.\ 2003,  Proc XXVIII ICRC
\re
6.\ Surdo A.\ 2003, Proc XXVIII ICRC

\endofpaper
\end{document}